\documentclass[doublespacing]{elsart}

\usepackage{graphicx}
\usepackage{amssymb}

\begin{document}
\begin{frontmatter}


\journal{SNS'2001: Version 1}


\title{Magnetic
Field Effects in the Pseudogap Phase: A Precursor Superconductivity
Scenario}

%
%
%
%
%
%
\author{Andrew P. Iyengar,}
\author{Ying-Jer Kao,}
\author{Qijin Chen,}
\author{K. Levin}

%
 
\address{}

%
%
%
%


%
%
%
%

\corauth[]{}


\begin{abstract}
We demonstrate
that the observed dependences of $T_c$ and $T^*$ on small magnetic fields
can be readily understood in a precursor
superconductivity approach to the pseudogap phase.
In this approach, the presence of a pseudogap at $T_c$ 
(but not at $T^*$) and the associated suppression of the density of states
lead to very different sensitivities to pair-breaking perturbations for 
the two temperatures.
Our semi-quantitative results address the puzzling experimental observation
that the coherence length $\xi$ is weakly dependent on hole
concentration $x$ throughout most of the
phase diagram. 
We present our results in a form which can be compared with the
recent experiments of Shibauchi et al, and argue that orbital effects
contribute in an important way to the $H$ dependence of $T^*$.
\end{abstract}

%



\end{frontmatter}

%
%
%
%
%


In the pseudogap phase, there are pronounced differences between the 
behavior of
the pseudogap onset temperature $T^*$ and the 
superconducting transition temperature $T_c$ with respect to magnetic 
fields.
In the underdoped regime,
$T_c$ is far more field sensitive than is $T^*$, and it has been 
argued that this lends support for the notion that the pseudogap
is unrelated to the superconductivity.  Contrary to this inference,
we demonstrate here that
these field dependences can be readily explained as a direct consequence of the 
pseudogap in a precursor superconductivity approach.
In this approach, the presence of a pseudogap at $T_c$ 
(but not at $T^*$) and the associated suppression of the density of states
lead to very different sensitivities to pair-breaking perturbations for 
the two temperatures.
Another effect of the pseudogap is that 
the coherence length $\xi$ does not necessarily 
coincide with other length scales
in the system, as occurs in BCS theory. 
In this paper, we illustrate these pseudogap effects at $T_c$ from the standpoint of 
Landau-Ginsberg theory and give a comparison of semi-quantitative 
results with recent experiments.

The various approaches to understanding the pseudogap phase of 
the high temperature
superconductors seem to be divided roughly into two schools: those
in which the pseudogap is associated with a competing energy gap
(or hidden order parameter\cite{Laughlin}) and those in which
the pseudogap derives from the superconductivity itself.  This
latter ``precursor superconductivity" school has multiple interpretations 
as well. The phase fluctuation\cite{Emery} and the spin-charge
separation schools\cite{Lee} are to be distinguished from the
present scheme in which the pseudogap arises in a mean-field generalization 
of BCS theory that allows the consideration of a strong pairing attraction. 
In contrast, the phase fluctuation school builds on \emph{strict} BCS
theory, but adds fluctuation effects which we neglect here.
Our approach is often referred to as a BCS-Bose Einstein crossover 
scenario, since a sufficiently strong attractive interaction allows
pairs to form at a temperature $T^*$ which is higher than 
the $T_c$ at which they Bose condense.
In the present paper, we use a formalism for treating the BCS-BEC crossover 
which we have extensively discussed and developed
previously\cite{Chen1,Chen2,Chen3}.

A fundamental feature of the crossover scenario is that 
increasing the attractive coupling strength $g$ introduces
bosonic as well as fermionic excitations. The presence of nonzero-momentum
bosonic pair excitations is responsible for both the pseudogap that develops
for temperatures $T_c < T < T^*$
as well as for the fact that below $T_c$, the excitation gap $\Delta$
is distinct from the superconducting order parameter $\Delta_{sc}$.  
The dispersion relation for fermions below $T_c$ has the BCS form with the 
full excitation gap $\Delta$ given by 
\begin{equation}
\Delta ^2 = \Delta_{sc} ^2 + \Delta_{pg}^2
\end{equation}
We will not give the self-consistency condition
on $\Delta_{pg}(T) $ here.
Notably, $\Delta_{pg}$ vanishes at zero temperature
(yielding $\Delta = \Delta_{sc}$) due to the total Bose condensation of all pairs.  
From this point of view, BCS theory is 
a weak coupling limit in which there are no non-zero-momentum pair excitations
and consequently $T^*$ coincides with $T_c$
and $\Delta = \Delta_{sc}$ at all temperatures below $T_c$.

We associate over- and under- doping with small
and large normalized coupling constants, respectively.
In all our calculations the coupling $g$ enters
in a dimensionless ratio with the bandwidth.  It is presumed
that as the Mott insulator is approached, the characteristic electronic
energy scales decrease-- so that even if $g$ is relatively $x$-independent,
its effectiveness increases with underdoping. 
The existence of a pseudogap above $T_c$ which develops as the coupling is 
increased differentiates
the physics of the underdoped cuprates from that of the overdoped
(BCS) state. 

We characterize the field sensitivity of $T_c$ through the coherence length $\xi$
defined by 
\begin{equation}
- \frac{1}{T_c} \left. \frac{d T_c}{dH} \right|_{H=0} 
= \frac{2\pi}{\Phi_0} \xi^2
\end{equation}
Similarly, we define a length scale $\xi^*$ for $T^*$.
In the case 
of $T_c$, where there is a true phase transition, we consider the 
free energy density of a linearized Landau-Ginsberg theory near $T_c$ :   
\begin{equation}
F = \left[ - \tau \left( 1 - \frac{T}{T_c} \right)
+ \eta ^ 2 (-i\nabla - 2\frac{e}{c} A )^2 \right]  | \Delta_{sc}| ^2
\end{equation}
The stiffness of the superconducting order parameter with respect to spatial variations
is characterized by $\eta^2$, whereas $\tau$ essentially measures the density of states
at the Fermi surface.   
As in BCS theory, $\xi^2 = \eta^2 / \tau$, although we must now consider the effect of the 
pseudogap on $\eta^2$ and $\tau$. 
In our formalism, the calculation of the (coupling dependent) parameters $\eta^2$ and $\tau$
is based on the application of the semi-classical
phase approximation to evaluate the pairing susceptibility in a small magnetic
field H \cite{LeePayne}; the details are given in \cite{us2}.
Although there is no phase transition at $T^*$, the pairing susceptibility 
provides natural analogs $\eta^{*2}$, $\tau^*$ with $\xi^{*2} = \eta^{*2} / \tau^*$.
In BCS theory, $\tau$ merely cancels the density of states appearing in $\eta^2$, resulting 
in a coherence length $\xi = C \cdot v_F/T_c$, where $C$ is a universal constant \cite{new}.
  In the pseudogap phase, this cancellation no longer occurs, and $\tau$ 
plays a more interesting role in determining $\xi$. In general, we find
that
$\xi$ is different
from
$C \cdot v_F/T_c$.

The doping dependence of $\xi$ and $\xi^*$ can be understood first from the fact
that the spatial stiffness $\eta^2$ (as well as $\eta^{*2}$) decreases with underdoping
because the stonger pairing interaction reduces the pair size \cite{Yamada}.
Second, $\tau$ and $\tau^*$ measure
the density of states at the Fermi energy at the respective temperatures.
The pseudogap present at $T_c$ (but not $T^*$) suppresses the density of 
states, causing $\tau$ to decrease rapidly with underdoping relative to $\tau^*$.
There is no analogous suppression of $\eta$ relative to $\eta^*$, allowing $\xi$ to 
differ from $\xi^*$ in the pseudogap phase. We find that 
$\xi^*$ decreases with underdoping, while the competition 
between the spatial stiffness and the density of states at $T_c$ results in a 
broad doping range over which $\xi$ is relatively constant, with 
a dramatic increase in $\xi$ as the superconductor-insulator boundary is approached. 
Quantitative results for the behavior of $\xi$, $\xi^*$ as functions
of hole concentration $x$ are given elsewhere\cite{us2}.

\begin{figure}
\centerline{\includegraphics[angle=0,width=6in,clip]{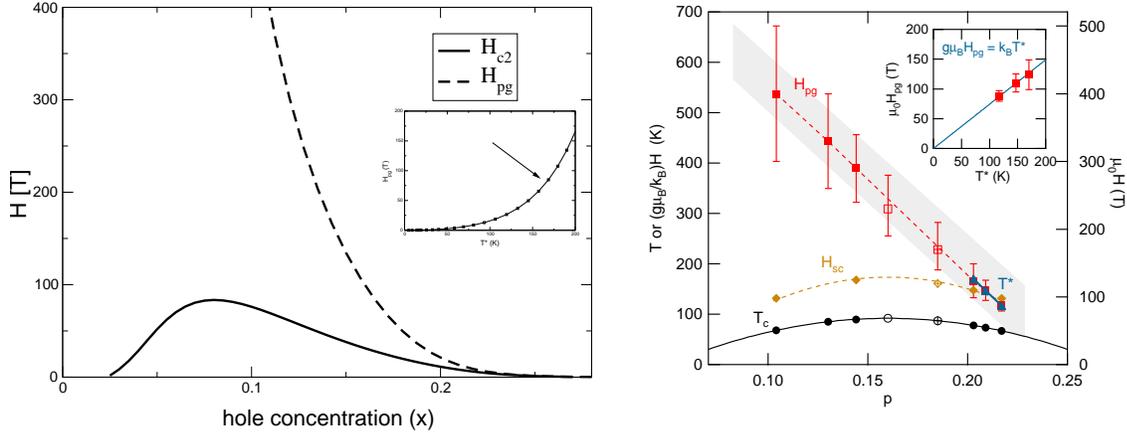}}
\caption{Magnetic fields required at $T=0$ 
to destroy superconductivity ($H_{sc}$ or $H_{c2}$)
and the pseudogap ($H_{pg}$) as extrapolated from (left) a calculation of
$\xi$, $\xi^*$ and 
(right) experiment \cite{Shibauchi}, plotted vs. hole concentration.
Insets plot $H_{pg}$ vs. $T^*$.
}
\label{fig:Shibauchi}
\end{figure}

In this paper we concentrate on an alternative representation of
these calculations, defining 
characteristic magnetic fields 
$H_{sc} = \Phi_0 / 2\pi\xi^2$ and 
$H_{pg} = \Phi_0 / 2\pi\xi^{*2}$.
By assuming straight-line extrapolations
on the $H-T$ phase diagram, we  
may interpret these as the fields required at zero temperature to destroy 
superconductivity and close the pseudogap, respectively.
This allows us to compare with
the recent study of BSSCO 2212 by Shibauchi et al \cite{Shibauchi}.
Figure \ref{fig:Shibauchi} demonstrates that our calculation (left)
captures the convergence of $H_{pg}$ and $H_{sc}$ in the overdoped region as
well as the divergence of these fields on the underdoped side as a consequence
of the opening of the pseudogap at $T_c$. 
The figure indicates that on the overdoped side, $T_c$ and $T^*$ 
are both sensitive to magnetic fields, whereas $T^*$ is far less field sensitive than 
$T_c$ on the underdoped side. 
We plot $H_{pg}$ versus $T^*$ (left, inset,) 
and in this way we can compare the apparent ``Zeeman scaling'' 
reported by Shibauchi et al\cite{Shibauchi} ($g \mu_0 H_{pg} = k_B T^*$)
with our calculations, which include only orbital magnetic effects.
(One might expect that as the temperature is lowered, and because the
appropriate critical fields become stronger, Zeeman coupling effects
will also need to be ultimately incorporated into a more complete 
theoretical treatment). 
However, in the doping region corresponding to the three data points
shown, there is reasonable agreement between the two
insets. More experimental data points at lower $T^*$
will ultimately be needed
to determine the relative importance of orbital and Zeeman coupling
in the magnetic field related pseudogap energy scales.

In conclusion, we have demonstrated that a precursor
pairing scenario is associated with
very different magnetic field sensitivities for $T^*$ and $T_c$. 
This
observation is contrary to the widely held belief that 
if the pseudogap is associated with superconducting pairing, then
$T^*$ and $T_c$ should be similarly dependent on the 
magnetic field strength.
Here we associate the different $H$ dependences with the fact
that a pseudogap is present at $T_c$ and absent at $T^*$.
The former observation leads to a modification of the Landau
Ginsberg coefficients
from their BCS
counterparts.
Moreover, the calculated field dependences of the two different
energy scales ($T^*$ and $T_c$)
appear to be in reasonable accord with experimental data.

%
%
%

%
%
%
%


\end{document}